\documentclass[12pt]{article}
\usepackage{amsmath}
\usepackage{graphicx}
\usepackage{enumerate}
\usepackage{natbib}
\usepackage{url} 

\usepackage{amssymb,physics}
\usepackage{mathrsfs,bm}
\usepackage[svgnames]{xcolor}
\usepackage{enumitem}
\usepackage[ruled]{algorithm2e}

\usepackage{amsthm}
\newtheorem{theorem}{Theorem}

\newtheorem{proposition}{Proposition}

\theoremstyle{remark}
\newtheorem{remark}{Remark}
\theoremstyle{definition}

\usepackage{xr-hyper}
\usepackage[bookmarksnumbered=true, colorlinks,
	citecolor=ForestGreen,       
	linkcolor=Blue,		         
	urlcolor=Magenta,	     	 
]{hyperref}
\makeatletter
\newcommand*{\addFileDependency}[1]{
\typeout{(#1)}
\@addtofilelist{#1}
\IfFileExists{#1}{}{\typeout{No file #1.}}
}
\makeatother
\newcommand*{\myexternaldocument}[1]{%
\externaldocument{#1}%
\addFileDependency{#1.tex}%
\addFileDependency{#1.aux}%
}
\myexternaldocument{supp-JASA}

\graphicspath{{./art/}}

\renewcommand{\Pr}{\mathbb{P}}
\newcommand{\E}{\mathbb{E}}

\begin{document}

\def\spacingset#1{\renewcommand{\baselinestretch}%
{#1}\small\normalsize} \spacingset{1}

\newcommand{\blind}{0}

\if0\blind
{
  \title{\bf Physics-encoded Spatio-temporal Regression}
  \author{Tongyu Li and Fang Yao\thanks{
    Fang Yao is the corresponding author: \texttt{fyao@math.pku.edu.cn}. This research is partially supported by  the National Key R\&D Program of China (No. 2022YFA1003801), the National Natural Science Foundation of China (No. 12292981, 11931001), the LMAM and the Fundamental Research Funds for the Central Universities, Peking University (LMEQF).}\\
    Department of Probability \& Statistics, School of Mathematical Sciences,\\ Center for Statistical Science, Peking University}
    \date{}
  \maketitle
} \fi

\if1\blind
{
  \bigskip
  \bigskip
  \bigskip
  \begin{center}
    {\LARGE\bf Physics-encoded Spatio-temporal Regression}
\end{center}
  \medskip
} \fi

\bigskip
\begin{abstract}
Physics-informed methods have gained a great success in analyzing data with partial differential equation (PDE) constraints, which are ubiquitous when modeling dynamical systems. 
Different from the common penalty-based approach, this work promotes adherence to the underlying physical mechanism that facilitates statistical procedures. 
The motivating application concerns modeling fluorescence recovery after photobleaching, which is used for characterization of diffusion processes. 
We propose a physics-encoded regression model for handling spatio-temporally distributed data, which enables principled interpretability, parsimonious computation and efficient estimation by exploiting the structure of solutions of a governing evolution equation. 
The rate of convergence attaining the minimax optimality is theoretically demonstrated, generalizing the result obtained for the spatial regression. 
We conduct simulation studies to assess the performance of our proposed estimator and illustrate its usage in the aforementioned real data example. 
\end{abstract}

\noindent%
\vfill

\newpage
\spacingset{1.9} 

\section{Introduction}\label{sec:intro}
Data evolving across spacetime are ubiquitous in our daily experiences, from the movement of vehicles on streets to the changing weather patterns in the atmosphere. 
The structural complexity of such data often originates in the physical mechanism that governs their generation, posing a formidable obstacle for traditional statistical modeling techniques. 
The problem-specific physics is generally formalized as a partial differential equation (PDE) to display intricate spatio-temporal dependencies \citep[see, e.g.,][Chapter 8]{palacios2022modeling}, since PDEs are extensively adopted in science and engineering as a powerful tool to characterize natural laws. 
When dealing with data governed by such complex dynamics, it is important to make use of prior physical knowledge to achieve better performance. 
There has been a burgeoning literature on developing physics-informed methods, where incorporating physical constraints brings new opportunities for scientific inquiry and potential benefits for learning algorithms \citep{karniadakis2021physics,meng2022physics,hao2022physics,willard2022integrating,brunton2024promising,yu2024learning}. 

In the context of supervised learning, a series of recent works, called spatial regression with PDE regularization \citep{azzimonti2015blood,sangalli2021spatial,arnone2023analyzing}, address the integration of physical information into the analysis of spatially (or spatio-temporally) distributed data. 
By including the regularization term that involves a PDE governing the phenomenon under study, their regression methods allow to account for spatial variation in a flexible and appropriate manner, which broadens the line of research at the crossroad between spatial statistics \citep{cressie2015statistics} and functional data analysis \citep{ramsay2005functional}. 
This approach is naturally extended to handle spatio-temporal data by \citet{bernardi2017penalized,arnone2019modeling}, demonstrating significant improvements upon classical methods for spatial data analysis. 
Meanwhile, the engineering community has made progress from the point of view of model calibration. 
For example, \citet{chen2022apik,peli2022physics} proposed physics-based kriging frameworks to leverage the PDE information. 
While the calibration method may prioritize fine tuning to match observed data, its reliance on empirical adjustment is likely to incline less to physical laws than the regularization method, resulting in inadequate efficacy  of estimation. 

Taking a different approach to spatio-temporal regression, this paper promotes adherence to the mechanism of physics that underlies the observation. 
We emphasize the insight that data can be projected to a compact latent space, given certain PDE constraint on them. 
Benefiting from this projection, the learning capacity is promisingly enhanced, which facilitates both computational tractability and estimation accuracy. 
Rather than the penalty-based physics-informed learning, our recipe shares the same spirit with \citet{rao2023encoding,ren2023physr}, who employed coercive encoding of physical knowledge to design the network architecture for modeling spatio-temporal dynamical systems. 
The philosophy of building physical constraints constructively has a long history of success, making it possible to effectively accelerate scientific computing \citep{faroughi2022physics}. 
In this sense, we take advantage of the global spectral method, unlike \citet{sangalli2021spatial} who exploits the local finite-element method. 
The spectral method constitutes a significant part of the arsenal of numerical computing for PDEs due to its inherent hierarchical structure, connections to approximation theory, and favorable convergence properties \citep{canuto2007spectral,kopriva2009implementing,shen2011spectral,meuris2023machine}. 
A general procedure of the spectral method is expanding the solution of a PDE as a linear combination of basis functions and estimating the coefficients of the linear combination so that the underlying PDE is satisfied in an appropriate way. 
As a benign byproduct, our proposed spectral estimator enjoys a closed form and thus delivers high interpretability. 

\begin{figure}[!ht]
    \centering
    \includegraphics[width=0.9\linewidth]{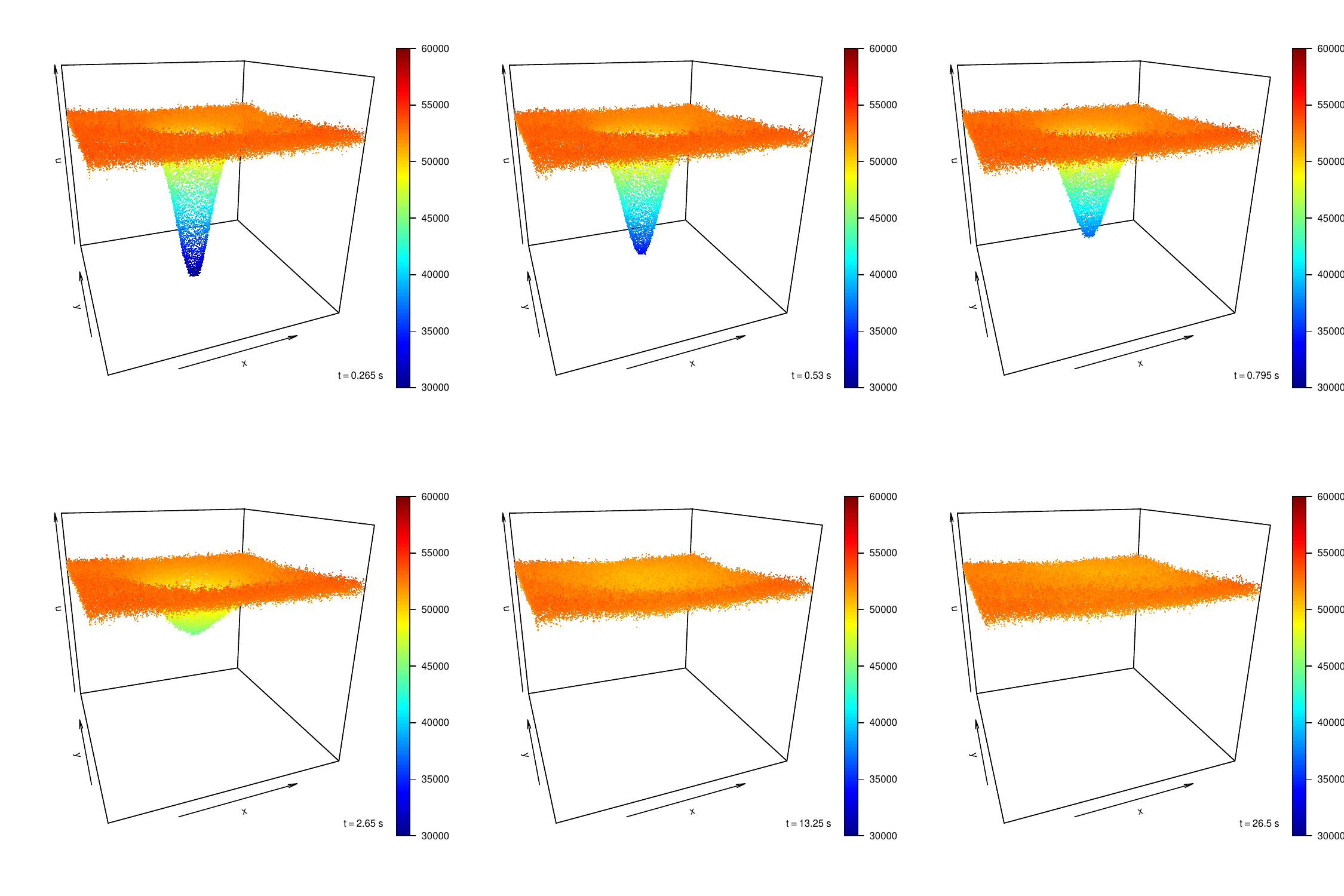}
    \caption{Spatio-temporal observations of fluorescence recovery after photobleaching. Plotted is the concentration of the fluorescent particles against the location coordinates associated with the bleach region at different post-bleach time points ($t = 0.265\operatorname{\mathrm{s}}, 0.53\operatorname{\mathrm{s}}, 0.795\operatorname{\mathrm{s}}, 2.65\operatorname{\mathrm{s}}, 13.25\operatorname{\mathrm{s}}, 26.5\operatorname{\mathrm{s}}$ from left to right, top to bottom).}
    \label{fig:frap_original}
\end{figure}

Our study is motivated from modeling fluorescence recovery after photobleaching \citep{moud2022fluorescence,waahlstrand2021deepfrap}, which is used for characterization of diffusion processes in materials science, pharmaceutics, food science and cell biology. 
As shown in Figure~\ref{fig:frap_original}, the time evolution of the concentration of the fluorescent particles is typically governed by a standard diffusion equation. 
Due to the explicit form of the differential operator, we can apply the proposed physics-encoded method to spatio-temporal regression in this setting, and the solution is firmly connected with the underlying mechanism. 

On the other hand, the regression model of interest falls into the large category of inverse problems, which are inextricably intertwined with statistics \citep{evans2002inverse,cavalier2008nonparametric,bissantz2008statistical}. 
Inverse problems for PDEs have been intensively studied by applied mathematicians; see \citet{hasanouglu2021introduction,isakov2017inverse} for a systematic introduction. 
To formulate the ill-posed nature of the inverse problem pertaining to evolution equations, there are mainly two types of data taken into consideration, namely, final value data \citep{muniz1999comparison,christensen2018final} 
and lateral Cauchy data \citep{klibanov2006estimates,li2020recovering}. 
Reaching a compromise on real observations, a few researchers get down to discrete data with random noise \citep{minh2018two,phuong2019cauchy}. 
Our proposed framework is a meaningful strategy in that interior points in spacetime, which play a crucial role in spatio-temporal regression, are now properly addressed. 
We mention that statistical inverse problems have also been investigated in an abstract way, that is, the unknown signal belongs to a readily manipulable Hilbert space \citep{bissantz2007convergence,loubes2008adaptive,nickl2020convergence}. 
In parallel, the Bayesian perspective has been increasingly studied recently \citep{knapik2011bayesian,nickl2023bayesian}, which made substantial progress on theoretical guarantees and inspired us to establish optimality for the proposed procedure.

The main contributions of this paper are twofold. 
First, we propose a new way of managing prior physical knowledge in statistical modeling, by introducing a physics-encoded regression model for tackling spatio-temporal data. 
Our method features obedience to the underlying evolution equation,  constructing estimators based on eigenmodes corresponding to the governing dynamical system so that the physical information can be  utilized. 
Second, we derive an upper bound on the estimation error under mild smoothness assumptions, as well as a matching lower bound in the minimax sense. 
To our best knowledge, this is the first consistency result for spatio-temporal regression with PDE constraints, which generalizes \citet{arnone2022some} that treated only spatial variation. 
It is worth mentioning that our rate of convergence is fundamentally different from the classical nonparametric rate which appears in the spatial case, since the data generated via an evolution equation carry inevitably an informational decay over time. 
The speculation by \citet{arnone2019modeling} is thus confirmed that the results obtained for regularized estimators cannot be directly extended to the spatio-temporal model considered here. 

The rest of this paper is organized as follows. 
Section~\ref{sec:method} begins with an elaborate exposition of model setting, and then presents the physics-encoded estimation method. 
In Section~\ref{sec:theoretical}, we establish theoretical results for the estimation error with matching upper and lower bounds. 
We examine the numerical performance of the proposed method via simulated and real data examples in Section~\ref{sec:numerical}. 
Concluding remarkds are offered in Section~\ref{sec:discuss}, while 
all technical proofs are deferred to an online Supplementary Material.

\section{Proposed Methodology}
\label{sec:method}
We consider a function $u : \mathcal{X}\times\mathcal{T} \to \mathbb{R}$ defined on the Cartesian product of a bounded spatial domain $\mathcal{X}\subset\mathbb{R}^{d}$ and a temporal interval $\mathcal{T} = [0,1]$. 
Suppose that the behavior of $u$ can be described in terms of an evolution equation 
\begin{equation}\label{eq:pde}
u' + \mathscr{L}u = 0 
\end{equation}
where $u'$ is the time derivative of $u$ and $\mathscr{L}$ is a linear differential operator taken with respect to the spatial variable. 
One of the frequently-encountered equations is the advection-diffusion equation \citep{bennett2013transport}, which arises often in transport phenomena associated with a turbulence degraded image \citep{carasso1978digital}, a contaminant plume \citep{skaggs1995recovering}, a ground heat exchanger \citep{larwa2018heat}, and so on. 

Given noisy measurements of $u$ from discrete locations in $\mathcal{X}\times\mathcal{T}$, a spatio-temporal regression model is of interest. 
Of the target function $u$ that solves \eqref{eq:pde}, there are $n$ measurements $(U_{i},X_{i},T_{i})$ with 
\begin{equation}\label{eq:obs}
U_{i} = u(X_{i},T_{i}) + \varepsilon_{i} ,\quad i=1,\dots,n,
\end{equation}
where $(X_{i},T_{i})$ is a sampling point and $\varepsilon_{i}$ represents an observational error. 
For the purpose of recovering $u$, we suggest a spectral method-based estimator that preserves the structure of solutions of \eqref{eq:pde}. 
Before outlining our estimation procedure, we provide a brief introduction to evolution equations for ease of exposition. 

We focus on the operator $\mathscr{L}$ that is time-invariant and elliptic, which implies an autonomous parabolic dynamical system. 
This covers most practical examples and can be generalized to time-dependent cases in the framework of spectral methods \citep{hesthaven2007spectral}. 
Then $u$ in \eqref{eq:pde} has the form \[ u(x,t) = \mathrm{e}^{-t\mathscr{L}}g_{0}(x) \] with $g_{0}=u(\bm{\cdot},0)$, allowing fast computation for $u$ especially in the case where $\mathscr{L}$ is diagonalizable. 
Even though $\mathscr{L}$ may fail to be self-adjoint, a remedy could be provided by introducing an offset operator using the Sturm--Liouville theory \citep{guenther2018sturm}. 
As an illustration, consider \[ \mathscr{L} = -\dv[2]{x} + 2p(x)\dv{x} + q(x) \] that acts on univariate functions. If $\mathscr{M}$ is the operator that multiplies a function by $\mathrm{e}^{\int_{0}^{x}p(\xi)\dd{\xi}}$, then 
\[ \mathscr{M}^{-1}\mathscr{L}\mathscr{M} = -\dv[2]{x} + \Big\{ q(x) + p^{2}(x) - \dv{p}{x}{(x)} \Big\} \] is self-adjoint. 
In view of this, suppose that there exists an invertible linear operator $\mathscr{M}$ such that $\tilde{\mathscr{L}} = \mathscr{M}^{-1}\mathscr{L}\mathscr{M}$ admits a spectral decomposition in terms of eigenvalue and eigenfunction pairs $(\lambda_{k},\phi_{k})$, $k=1,2,\dots$, where $\phi_{k}$'s are normalized so that their inner products in $L^2(\mathcal{X})$ are given by Kronecker's delta: 
\[ \int_{\mathcal{X}} \phi_{k}\phi_{k'} \dd{x} = \delta_{kk'} = \begin{cases}
1, \  \ \text{ if } k=k' ,\\
0, \ \  \text{ otherwise.}
\end{cases}\]
Denote $\psi_{k} = \mathscr{M}\phi_{k}$. It follows that $(\lambda_{k},\psi_{k})$, $k=1,2,\dots$, are eigenvalue and eigenfunction pairs of $\mathscr{L}$. 
Since $\mathscr{M}$ involves no temporal change, \eqref{eq:pde} can be converted to \[ v' + \tilde{\mathscr{L}}v = 0 \] with $v=\mathscr{M}^{-1}u$. 
Therefore, if the initial value is written as $g_{0} = \sum_{k=1}^{\infty}\alpha_{k}\psi_{k}$, then \[ u(x,t) = \sum_{k=1}^{\infty} \alpha_{k} \mathrm{e}^{-\lambda_{k}t} \psi_{k}(x) .\]
In this way, the estimation problem for $u$ on $\mathcal{X}\times\mathcal{T}$ boils down to that for $g_{0}$ on $\mathcal{X}$. 
The physical information of \eqref{eq:pde} is leveraged so that $u$ can be determined through $g_0$, which reduces the regression model \eqref{eq:obs} and facilitates subsequent statistical analysis. 

Given how the regression function evolves across time, the preceding argument indicates that it suffices to address the recovery of initial values. 
We now present a quantitative verification for corresponding approximation. 
Regarding the error that varies with time, we deduce an upper bound in Proposition~\ref{prop:evo} below. 
To this end, for any function $f : \mathcal{X} \to \mathbb{R}$, let $\norm{f}$ be the square root of 
\begin{equation*}
\norm{f}^{2} = \int_{\mathcal{X}} \abs{\mathscr{M}^{-1}f}^{2} \dd{x} .
\end{equation*}
Note that the above-defined $\norm{\bm{\cdot}}$ is equivalent to the $L^{2}(\mathcal{X})$-norm under mild conditions on $\mathscr{M}$, e.g., $C_{\mathscr{M}}^{-1}\abs{f} \le \abs{\mathscr{M}f} \le C_{\mathscr{M}}\abs{f}$ pointwise for some constant $C_{\mathscr{M}}>0$. 
\begin{proposition}\label{prop:evo}
Let $\hat{u}(x,t) = \sum_{k=1}^{\infty} \hat{\alpha}_{k} \mathrm{e}^{-\lambda_{k}t} \psi_{k}(x)$. For every $t\in\mathcal{T}$, \[ \norm{(\hat{u}-u)(\bm{\cdot},t)} \le \mathrm{e}^{-(\inf_{k}\lambda_{k})t} \norm{(\hat{u}-u)(\bm{\cdot},0)} .\]
\begin{remark}\label{rmk:eigen}
The $\lambda_{k}$'s are supposed to have a lower bound, which is a common entailment of ellipticity. 
Indeed, if $\tilde{\mathscr{L}}$ is elliptic of degree $m$, then a generalization of Weyl's theorem \citep[Theorem~1.3]{zielinski1998asymptotic} suggests that the asymptotic order of $\lambda_{k}$ is $k^{m/d}$. 
Besides, thanks to the flatness of $\mathcal{X}$, the eigenfunctions $\phi_{k}$, $k=1,2,\dots$, are uniformly bounded \citep{toth2002riemannian}. 
Therefore, the eigenmodes of $\tilde{\mathscr{L}}$, and of $\mathscr{L}$, are theoretically considered as similar to the eigenmodes of the typical Laplacian operator. 
\end{remark}
\end{proposition}

In order to estimate $g_{0}$ based on discrete observations from \eqref{eq:obs}, we adopt a least-squares criterion taking the knowledge of \eqref{eq:pde} into account. Write 
\begin{equation*}
u_{g}(x,t) = \mathrm{e}^{-t\mathscr{L}}g(x) = \mathscr{M}\mathrm{e}^{-t\tilde{\mathscr{L}}}\mathscr{M}^{-1}g(x)
\end{equation*}
so that $u = u_{g_{0}}$. 
Let the estimator $\hat{g}$ be the minimizer of \[ \sum_{i=1}^{n}\{ U_{i} - u_{g}(X_{i},T_{i}) \}^{2} \]
among \[ g = \sum_{k=1}^{K}a_{k}\psi_{k} \quad (a_{1},\dots,a_{K}\in\mathbb{R}) \] for some integer $K=K_{n}$. 
The cutoff $K$ serves as a tuning parameter that ensures computational feasibility and controls regularity. 
It can be easily obtained that $\hat{g} = \sum_{k=1}^{K}\hat{\alpha}_{k}\psi_{k}$ with 
\begin{equation}\label{eq:est}
(\hat{\alpha}_{k})_{1\le k\le K} = \Big(\sum_{i=1}^{n} Z_{iK}Z_{iK}^{\top} \Big)^{-1} \sum_{i=1}^{n} Z_{iK}U_{i} ,
\end{equation}
where $Z_{iK} = (\mathrm{e}^{-\lambda_{k}T_{i}}\psi_{k}(X_{i}))_{1\le k\le K}$. 
Following Proposition~\ref{prop:evo}, we set $\hat{u} = u_{\hat{g}}$ as an estimator for $u$, which again reflects the physical constraints. 
A reminder for practice is that the eigenvalue and eigenfunction pairs $(\lambda_{k},\psi_{k})$, $k=1,\dots,K$, can be effectively computed, e.g., using the method developed by \citet{platte2004computing}. 

This reduced-order modeling strategy leads to low-dimensional approximation and alleviates computational burden in that it explicitly appeals to the physical structure, whose viewpoint paves the way for a general framework of deriving structure-preserving surrogate models \citep{ghattas2021learning}.
We mention here how to handle the challenge of model reduction for nonlinear systems. 
Although we may approximate the state in a finite-dimensional subspace of functions, the nonlinear term could still require computations that scale with the full order. To remedy this and achieve a reduced model that is efficient to solve, a popular class of methods introduce an extra layer of approximation referred to as hyper-reduction, e.g., \citet{chaturantabut2010nonlinear} proposed the discrete empirical interpolation for the nonlinear term. 
In this way, the inverse problem associated with a nonlinear evolution equation can be dealt with using the aforementioned method, which facilitates computation while keeping the structure of solutions easily interpretable within suitable physical sense. 

As a notable competitor, \citet{arnone2019modeling} employed a penalty to suppress departures from the problem-specific PDE description. Then a finite-element basis is constructed to span the function space to be searched. 
In comparison, we explicitly incorporate physical knowledge to obtain an eigenfunction basis. Allowing for better insights into the underlying mechanism of the problem, our proposed framework is easier to implement and interpret. 
This is because the overall formulation becomes simpler and more straightforward by removing the penalty term and reorganizing the solution space, which also alleviates the computational burden. 
Finite-element methods involve mesh generation, which can be complex and computationally expensive, especially for problems with irregular geometries or adaptive mesh refinement. By contrast, the proposed spectral method do not require mesh generation since we operate in the spectral domain, which simplifies the preprocessing stage. 
Moreover, we handle boundary conditions more naturally compared to the penalty-based method. Since global basis functions are used, our estimator inherently satisfies boundary conditions at all points in the domain. 
The spectral nature of our method also provides higher estimation accuracy than the penalty-based method, especially for smooth solutions with low noise that support fast convergence, which will be demonstrated later in simulation studies. 
While the proposed method excels in problems with smooth solutions and regular geometries, it may face challenges in handling discontinuities or singularities. Penalized criteria combined with finite-element methods, however, are more versatile and suitable for a wider range of problems through appropriate strategies of penalty design and mesh refinement. 
The choice between the two methods ultimately depends on the specific characteristics of the problem under study and the computational resources available.

\section{Theoretical Guarantees}
\label{sec:theoretical}
In this section, we establish the minimax convergence rate of the proposed estimator $\hat{g}$ given in \eqref{eq:est}. 
The rate is optimal with respect to certain regularity conditions, which guarantees the efficiency of estimation. 
To begin with, we consider a random design that $(X_{i},T_{i},\varepsilon_{i})$ in \eqref{eq:obs} are independent and identically distributed as $(X,T,\varepsilon)$. The random variable $\varepsilon$ is supposed to have zero mean and finite variance and is independent of $(X,T)$, which is standard in the context of regression. 
We now impose some assumptions on the distribution of $(X,T)$ associated with the underlying evolution equation. 
In what follows, denote $A \lesssim B$ or $B \gtrsim A$ for real quantities $A$ and $B$ when there exists some constant $C>0$ such that $A \le CB$. Moreover, $A \asymp B$ means $A \lesssim B$ and $A \gtrsim B$ simultaneously. 
\begin{enumerate}[label=({A}\arabic*)]
    \item\label{asm:cov} There exists some constant $c\ge 0$ such that for any $K$, \[ \E\bigg[\Big\{\sum_{k\le K} \mathrm{e}^{-2\lambda_{k}T} \psi_{k}^{2}(X)\Big\}^{2}\bigg] \lesssim K^{c} .\]
    \item\label{asm:eigen} If $Z_{K} = (\mathrm{e}^{-\lambda_{k}T}\psi_{k}(X))_{1\le k\le K}$, then the smallest eigenvalue of $\E(Z_{K}Z_{K}^{\top}) \in \mathbb{R}^{K\times K}$ is \[ \nu_{K} \gtrsim K^{-r} \] for any $K$ and some constant $r\ge 1 - c/2$.
    \item\label{asm:coef} There exists some constant $s>(r+c)/2$ such that for any $K$, \[ \sum_{k>K}\alpha_{k}^{2} \lesssim K^{-(2s-1)} ,\] \[ \E\bigg[\Big\{\sum_{k>K}\alpha_{k}\mathrm{e}^{-\lambda_{k}T}\psi_{k}(X)\Big\}^{2}\bigg] \lesssim K^{-(r+2s-1)} .\]
\end{enumerate}

The assumption~\ref{asm:cov} implies that the working covariates in $Z_{K}$ are square-summable in the sense of second moment, and hence requires $r\ge 1 - c/2$ in assumption~\ref{asm:eigen} as \[ K\nu_{K} \le \tr\{\E(Z_{K}Z_{K}^{\top})\} = \E\{\tr(Z_{K}Z_{K}^{\top})\} \lesssim K^{c/2} .\]
If $\E\{\mathrm{e}^{-2(\lambda_{k_1}+\lambda_{k_2})T}\} \lesssim \max(k_{1},k_{2})^{-r}$ and $\psi_{k}$'s are uniformly bounded, as indicated by Remark~\ref{rmk:eigen}, then assumption~\ref{asm:cov} holds with $c=0$ if $r>2$; $c=2-r$ if $r<2$; and any $c>0$ if $r=2$. 
The assumption~\ref{asm:eigen} prevents the singularity of design matrices, whose presence stands out in the choice of $K$. 
The assumption~\ref{asm:coef} determines the smoothness class of regression functions which can be aligned relative to the covariates. In nonparametric statistics, it is important to utilize smooth functions so that the approximation error is conveniently reduced \citep{tsybakov2009introduction}. 
Simple calculations lead to assumptions \ref{asm:eigen} and \ref{asm:coef} in the typical case where $X$ and $T$ are independent, $\E\{\psi_{k}(X)\psi_{k'}(X)\} = \delta_{kk'}$, $\E(\mathrm{e}^{-2\lambda_{k}T}) \asymp k^{-r}$ and $\abs{\alpha_{k}} \lesssim k^{-s}$. 

Thanks to the explicit closed-form solution in \eqref{eq:est}, the consistency of our proposed estimator can be established. 
Bearing in mind the trade-off between bias and variance, we derive in Theorem~\ref{thm:rate} the estimation error with respect to the sample size and the value of the smoothing parameter. 
\begin{theorem}\label{thm:rate}
Under assumptions \ref{asm:cov}--\ref{asm:coef}, \[ \norm{\hat{g}-g_{0}}^{2} = \mathcal{O}_{\Pr}\{K^{-(2s-1)}+n^{-1}K^{1+r}\} .\]
In particular, if $K\asymp n^{1/(r+2s)}$, then \[ \norm{\hat{g}-g_{0}}^{2} = \mathcal{O}_{\Pr}\{n^{-(2s-1)/(r+2s)}\} .\]
\end{theorem}
The convergence rate $n^{-(2s-1)/(r+2s)}$ is achieved with an appropriately chosen cutoff $K$, which will be demonstrated to attain the minimax optimality in Theorem~\ref{thm:minimax} below. 
As an immediate consequence, we have \[ \sup_{t\in\mathcal{T}} \norm{(\hat{u}-u)(\bm{\cdot},t)}^{2} = \mathcal{O}_{\Pr}\{n^{-(2s-1)/(r+2s)}\} \] by Proposition~\ref{prop:evo}. 
It is intriguing to see how the rate $n^{-(2s-1)/(r+2s)}$ is connected with the classical nonparametric rate $n^{-2r/(2r+1)}$, where $r=m/d$, as discussed in Remark~\ref{rmk:eigen}, keeps its meaning in assumption~\ref{asm:eigen}. 
To this end, notice that \[ \norm{\mathscr{L}g_0}^2 = \sum_{k=1}^{\infty} (\lambda_k \alpha_k)^2 \asymp \sum_{k=1}^{\infty} (k^r \alpha_k)^2 .\]
In order to ensure that $\norm{\mathscr{L}g_0} < \infty$ in the case where $\limsup k^s \abs{\alpha_k} > 0$, we are supposed to require $2(r-s)<-1$, or equivalently, $s > r + 1/2$. 
The extreme point $s = r + 1/2$ plugged into $n^{-(2s-1)/(r+2s)}$ yields $n^{-2r/(3r+1)}$, which is slower than $n^{-2r/(2r+1)}$ when approaching zero. 
This is not so surprising because we estimate $g_0$ using observations drawn from various time points rather than exactly at a fixed instant. While the noise level associated with measurement is constant, the information of $g_0$ decays over time $t$ with discount factor $\mathrm{e}^{-t\mathscr{L}}$. This makes the denominator $2r+1$ increase by $r$, and thus the convergence slows down. 

It is pointed out that convergence rates of the form $n^{-(2s-1)/(r+2s)}$ are generic to a large class of inverse problems where the difficulty of inverting the operator increases with $r$ and the smoothness of the target function increases with $s$ \citep{hall2007methodology}. 
Our result complements the literature that highlights this nature, which indicates a powerful framework of validating consistency for estimation problems given discrete observations with certain PDE constraint. 

Then we show that the convergence rate in Theorem~\ref{thm:rate} is in fact optimal by providing a minimax lower bound. 
Denote by $P$ the distribution of $(X,T,\varepsilon)$. The model \eqref{eq:obs} can be specified now by $(g_{0},\mathscr{L},P)$. Let $\mathcal{P}$ be the collection of $(g_{0},\mathscr{L},P)$ that satisfy assumptions \ref{asm:cov}--\ref{asm:coef}. 
The minimax lower bound is given in the following theorem.
\begin{theorem}\label{thm:minimax}
There exists a constant $\delta > 0$ such that 
\[ \liminf_{n\to\infty} \inf_{\tilde{g}} \sup_{(g_{0},\mathscr{L},P)\in\mathcal{P}} \Pr_{(g_{0},\mathscr{L},P)}\{ \norm{\tilde{g}-g_{0}}^{2} > \delta n^{-(2s-1)/(r+2s)} \} > 0 ,\]
where $\inf_{\tilde{g}}$ is taken over all possible estimators $\tilde{g}$ based on observations from \eqref{eq:obs}. 
\end{theorem}

We end this section with a brief technical note on the case of a fixed design where points of measurement are pre-specified. 
In parallel with Theorems~\ref{thm:rate} and \ref{thm:minimax}, similar results are also derivable when the assumptions are appropriately modified. For example, the population means appearing in assumptions \ref{asm:cov}--\ref{asm:coef} may be replaced by the sample means since observations are deterministic rather than random. The proof could proceed by analogy and thus is omitted for conciseness.

\section{Numerical Studies}
\label{sec:numerical}

\subsection{Simulation}
We take for instance the one-dimensional diffusion equation with Neumann boundary conditions to demonstrate the usefulness of our method. 
The spatial-temporal domain is $\mathcal{X}\times\mathcal{T} = [0,1]\times[0,1]$ and the differential operator $\mathscr{L}$ is the negative Laplacian $-\Delta$. Then the eigenvalue and eigenfunction pairs are analytically given by $(\lambda_1,\psi_1(x)) = (0,1)$ and $(\lambda_{k+1},\psi_{k+1}(x)) = ((k\pi)^2, 2^{1/2}\cos(k\pi x))$ for $k=1,2,\dots$. It follows that the smoothness of $\mathscr{L}$ is measured by $r=2$. 
The initial values are specified as $g_0 = \sum_{k=1}^{50} \alpha_k \psi_k$ with $\alpha_1 = 0.3$ and $\alpha_k = 4 (-1)^{k-1} k^{-s}$ for $k \ge 2$, where $s = 2$. 
We generate a sample of $n = 200$ subjects according to \eqref{eq:obs}, where the spatial-temporal location $(X_i,T_i)$ is uniformly sampled and the noise $\varepsilon_i$ is randomly drawn from the normal distribution $\mathcal{N}(0,\sigma^2)$ with $\sigma = 0.2$. 
To select the tuning parameter $K$, we minimize the Bayesian information criterion, 
\begin{equation}\label{eq:BIC}
\textsc{bic} = n \log(\textsc{rss}/n) + \log(n) K ,
\end{equation}
where \textsc{rss} is the squared sum of residuals $\sum_{i=1}^{n}\{U_i-u_{\hat{g}}(X_i,T_i)\}^2$. 
The quality of our estimator is assessed by the integrated squared error of the initial value function, $\textsc{ise} = \norm{\hat{g}-g_{0}}^{2}$. 
Under different choices of $K$, the results are reported in Table~\ref{tab:select}. It can be seen that the selection of the tuning parameter is valid. Moreover, the spectral nature of the proposed estimator leads to a sufficiently small number of parameters to be computed. This not only reduces computation burden but also enhance the interpretability, where each component $\hat{\alpha}_k \psi_k$ has an evolutionary pattern determined merely by $\lambda_k$. 
\begin{table}[!ht]
    \centering
    \caption{\textsc{ise} and \textsc{bic} under different cutoffs. Reported are the average and standard deviation (in the parenthesis) based on 200 Monte Carlo replications.}
    \label{tab:select}
\begin{tabular}{crrrrr}
$K$ & 1~ & 2~ & 3~ & 4~ & 5~ \\ \hline
\textsc{ise} & 1.318~ & 0.327~ & 0.159~ & 0.373~ & 1.339~ \\
 & ($<$0.001) & ($<$0.001) & ($<$0.001) & (0.008) & (0.015) \\ \hline
\textsc{bic} & $-471.0$~ & $-623.0$~ & $-630.4$~ & $-627.5$~ & $-623.3$~ \\
 & (0.2) & (0.1) & (0.1) & (0.1) & (0.1) \\ \hline
\end{tabular}
\end{table}

\begin{figure}[!ht]
    \centering
    \includegraphics[width=0.48\linewidth]{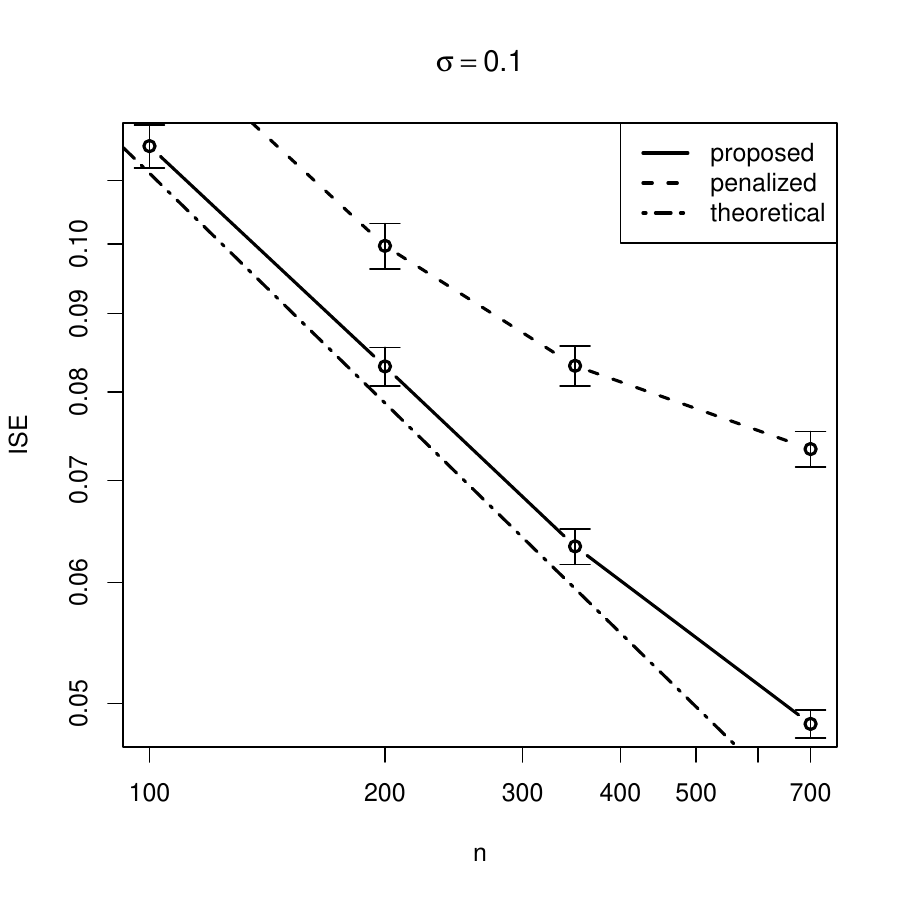}
    \quad
    \includegraphics[width=0.48\linewidth]{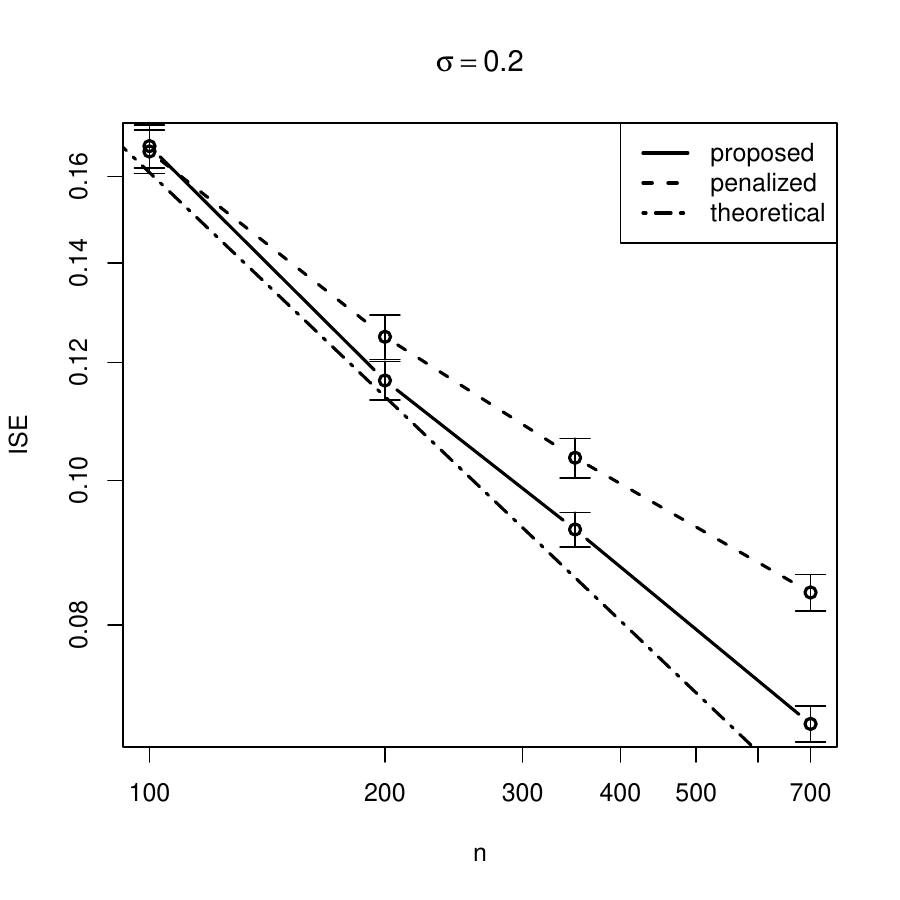}
\\
    \includegraphics[width=0.48\linewidth]{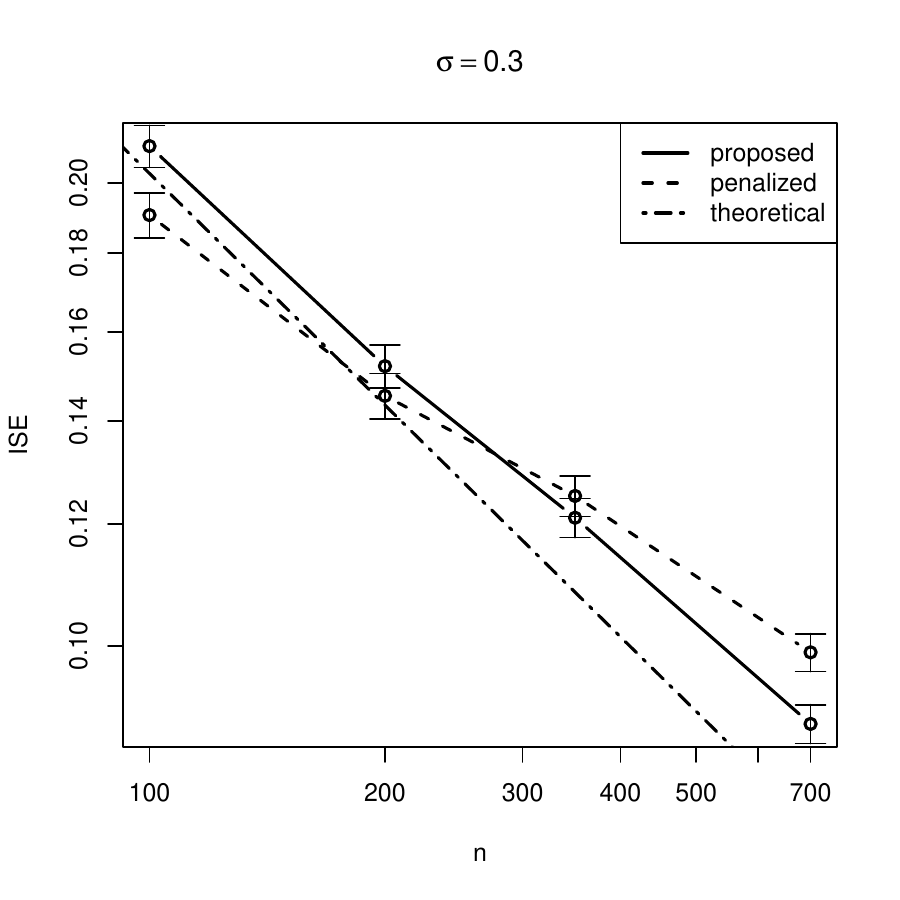}
    \quad
    \includegraphics[width=0.48\linewidth]{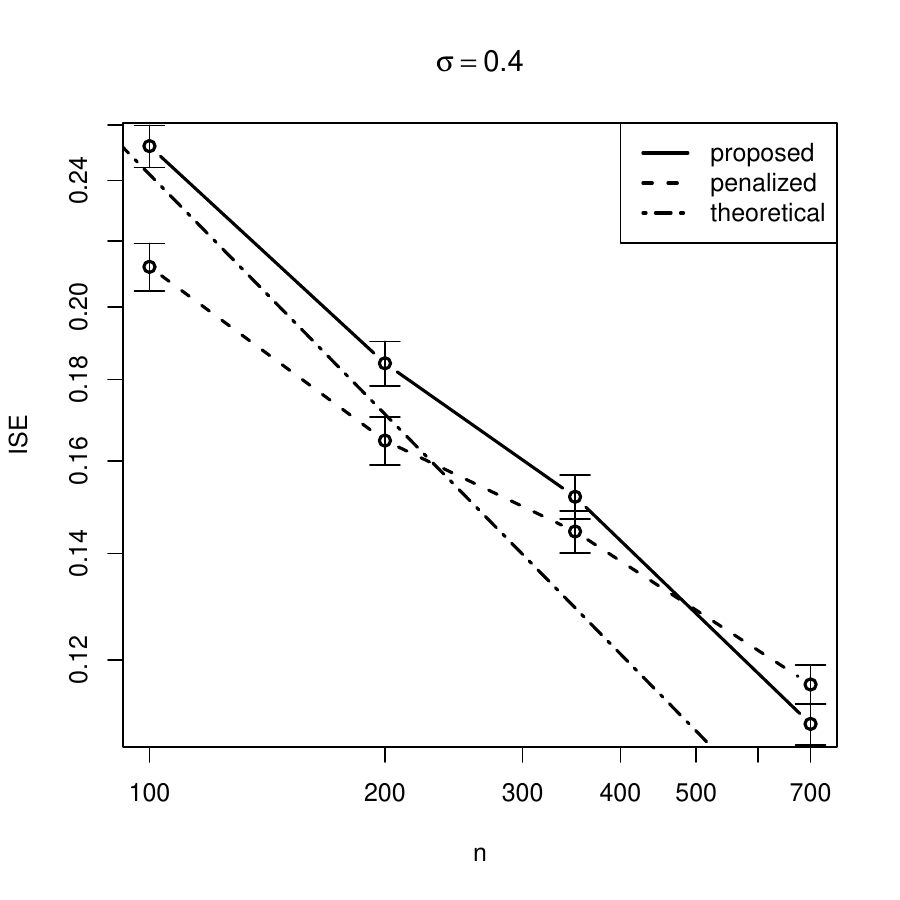}
    \caption{\textsc{ise} versus the sample size $n$ under different noise level $\sigma$. Both the proposed estimate and the regularization estimate were computed with tuning parameter chosen to minimize the \textsc{ise}. The circle and error bars correspond to average and average $\pm$ standard deviation, respectively, all based on 200 Monte Carlo replications. Both axes are in log scale to yield better visualization.}
    \label{fig:simu}
\end{figure}

To illustrate the efficiency of the proposed method, we further conduct simulation studies with varying sample sizes $n$ and noise levels $\sigma$. 
For comparison with \citet{arnone2019modeling} where sampling designs are aligned in spacetime, we construct a spline estimator with regularization that incorporates the underlying PDE and boundary conditions by adding penalty terms to the square loss. For computational convenience, the finite element method is not bothered with, as the currently considered spatial-temporal domain $\mathcal{X}\times\mathcal{T}$ is regular. 
The \textsc{ise} of both estimates are shown in Figure~\ref{fig:simu}, together with the theoretical rate that is proportional to $n^{-(2s-1)/(r+2s)} = n^{-1/2}$. 
As the sample size increases, the consistency of estimation is empirically illustrated and the convergence rate of our proposed estimator is seen to coincide with the theoretical findings. For a sufficiently large sample, the proposed cutoff estimator tends to have a better performance than the estimator based on regularization. 
Besides, a smaller noise level could improve the estimation accuracy, where the gap between the two estimation methods is even more pronounced. 

\subsection{Real Data Example}
Here we revisit the data example of fluorescence recovery after photobleaching depicted in Fig.~\ref{fig:frap_original}, which results from experiments by \citet{waahlstrand2021deepfrap}. 
The raw dataset is downloaded from \url{https://zenodo.org/records/3874218}, where we take the mean of twenty samples of experimental data in sucrose-water solutions with 32 w/w\% sucrose. 
In an experiment, fluorescent particles are photobleached by a high-intensity laser in a bleach region. Unbleached particles will move into the bleach region, leading to a recovery of the fluorescence intensity. 
The evolution of the concentration $u$ of the fluorescent particles is described by the two-dimensional diffusion equation \[ u' - D \Delta u = 0 ,\] where $D$ is the diffusion coefficient depending on the experimental setting. 
It is known from previous studies that $D$ can be set as $8.9\times 10^{-11} \,\mathrm{m^2/s}$, and the spatial domain is a square with side length $1.945\times 10^{-4} \,\mathrm{m}$, which requires periodic boundary conditions on $u$. 
The measurement of the particle concentration $u$ is taken on a $256\times 256$ grid with pixel size $7.598 \times 10^{-7} \,\mathrm{m}$ for $100$ post-bleach frames, where the time lag between consecutive frames is $0.265 \,\mathrm{s}$, and thus the sample size is $n = 6553600$. 
Six selected post-bleach frames are displayed in Fig.~\ref{fig:frap_original}, reflecting the evolving behavior of $u$. The concentration profile becomes flat as time passes, which corresponds to the decay of information over time. 

\begin{figure}[!ht]
    \centering
    \includegraphics[width=0.9\linewidth]{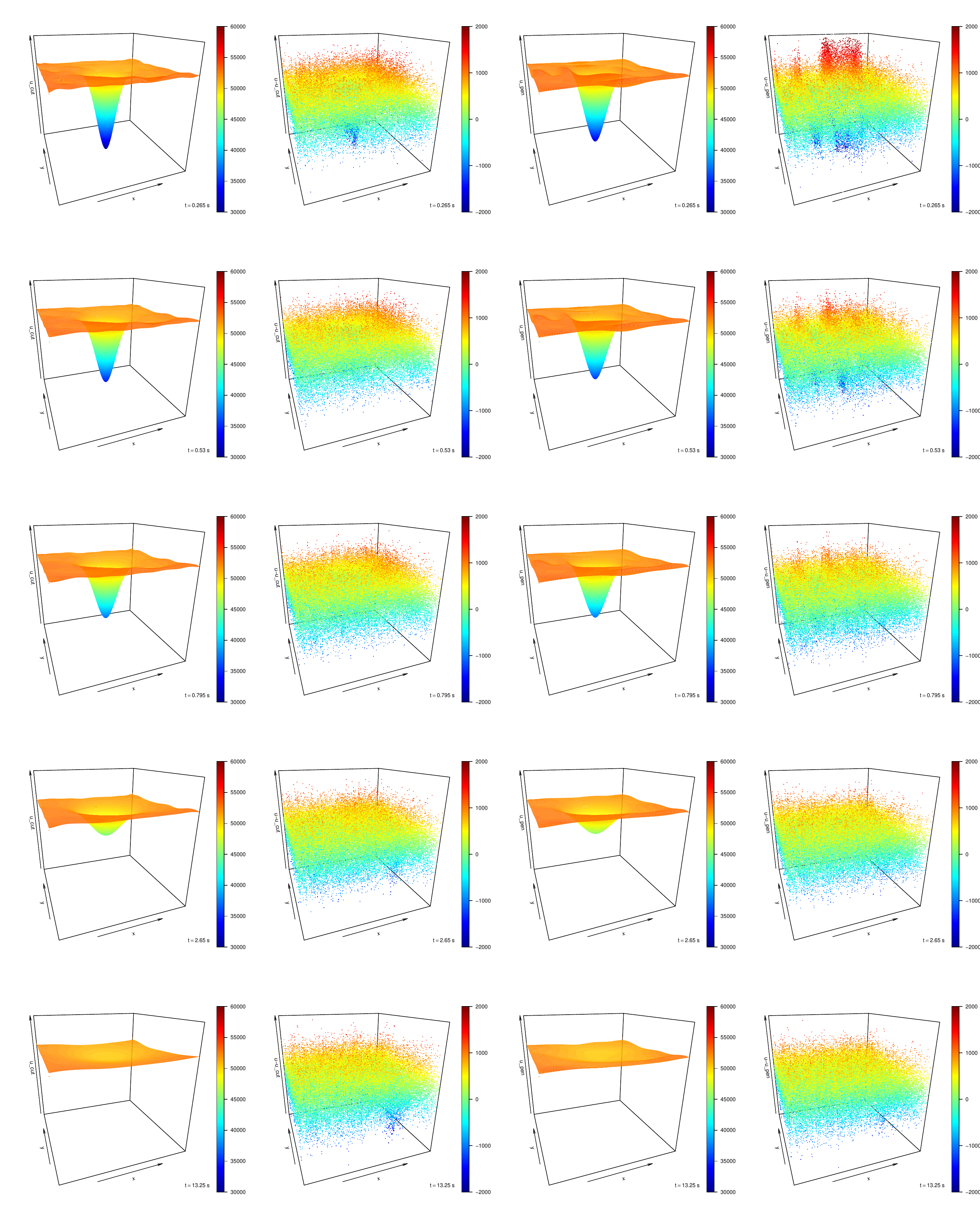}
    \caption{Estimated concentration of the fluorescent particles and residuals by the proposed method (Columns 1 and 2) and the regularization method (Columns 3 and 4) at different post-bleach time points ($t = 0.265\operatorname{\mathrm{s}}, 0.53\operatorname{\mathrm{s}}, 0.795\operatorname{\mathrm{s}}, 2.65\operatorname{\mathrm{s}}, 13.25\operatorname{\mathrm{s}}$ from top to bottom).}
    \label{fig:frap_fit}
\end{figure}

We apply the proposed method and the regularization method to the estimation of the particle concentration $u$. 
To determine the tuning parameter, the \textsc{bic} in \eqref{eq:BIC} is adopted for the proposed estimator and the generalized cross-validation is used for the regularization estimator. 
The estimation results are presented in Figure~\ref{fig:frap_fit} at several time points, which fit the data well since the residuals have a smaller order of magnitude than the original observed values. 
The estimation accuracy borrows strength from the underlying physical mechanism and the estimated concentration function may help people understand the diffusion phenomenon across time. 
In particular, the proposed method is more parsimonious than the regularization method, where the numbers of parameters to be computed are $19\times 20 = 380$ and $9^3 = 729$, respectively. 
It is notable that the starting values of the concentration $u$ are estimated more steadily by our method, giving rise to a milder landscape of residuals which implies more even dispersion. Since our primary goal is to recover the initial value function, it is not surprising that the proposed method outperforms the regularization estimator at the starting stage. Due to the correct specification of physical knowledge, it also meets the evolutionary trend and has a good overall performance. 

\section{Discussion}
\label{sec:discuss}
This work presents a method for spatial-temporal regression that takes into account the underlying PDE constraints by constructing estimators with prespecified structure of solutions. In particular, the estimation procedure becomes easy to implement and interpret. We demonstrate that the convergence rate of the proposed estimator achieves minimax optimality under mild assumptions, whose superiority is also illustrated via numerical examples as compared with the existing regularization method. It is believed that making thorough use of physical knowledge can lead to parsimonious and effective modeling, and deserves wider and deeper applications. 

In the proposed approach to nonparametric estimation, correct specification of underlying PDEs is indispensable. To broaden the applicability of our method, a beneficial modification is to learn the PDE structure from data. The current line of research addresses this issue in its own right, e.g., \citet{xun2013parameter} considered specifically parameter estimation in PDEs. A possible improvement is to derive a semiparametric framework that simultaneously handles the parametric part, which plays a dominating role in characterizing the properties of PDEs, and the nonparametric part, which presents the global pattern. The nonparametric part can be also regarded as a minor correction to the PDE such that efficient inference for the parametric part is still feasible. We will treat these extensions in future work.

\appendix

\bigskip
\begin{center}
{\large\bf SUPPLEMENTARY MATERIAL}
\end{center}

\begin{description}
\item[supp] Proofs for all theoretical results in this paper. (.pdf file)
\item[code] R-code to implement and reproduce the simulation and real data analysis together with corresponding raw datasets. 
\end{description}


\bibliographystyle{agsm}
\bibliography{ref}

@article{moud2022fluorescence,
  title={Fluorescence recovery after photobleaching in colloidal science: introduction and application},
  author={Moud, Aref Abbasi},
  journal={ACS Biomaterials Science \& Engineering},
  volume={8},
  number={3},
  pages={1028--1048},
  year={2022},
  publisher={ACS Publications}
}

@article{waahlstrand2021deepfrap,
  title={DeepFRAP: Fast fluorescence recovery after photobleaching data analysis using deep neural networks},
  author={W{\aa}hlstrand Sk{\"a}rstr{\"o}m, Victor and Krona, Annika and Lor{\'e}n, Niklas and R{\"o}ding, Magnus},
  journal={Journal of microscopy},
  volume={282},
  number={2},
  pages={146--161},
  year={2021},
  publisher={Wiley Online Library}
}

@book{tsybakov2009introduction,
  author="Tsybakov, Alexandre B.",
  title="Introduction to nonparametric estimation",
  year="2009",
  publisher="Springer New York"
}

@book{wainwright2019high,
  title={High-dimensional statistics: A non-asymptotic viewpoint},
  author={Wainwright, Martin J},
  volume={48},
  year={2019},
  publisher={Cambridge university press}
}

@book{kallenberg2021foundations,
  author="Kallenberg, Olav",
  title="Foundations of modern probability",
  year="2021",
  volume={99},
  publisher="Springer International Publishing"
}

@book{guenther2018sturm,
  title={Sturm-Liouville Problems: Theory and Numerical Implementation},
  author={Guenther, Ronald B and Lee, John W},
  year={2018},
  publisher={CRC Press}
}

@article{zielinski1998asymptotic,
  title={Asymptotic distribution of eigenvalues for some elliptic operators with simple remainder estimates},
  author={Zielinski, Lech},
  journal={Journal of Operator Theory},
  pages={249--282},
  year={1998},
  publisher={JSTOR}
}

@article{toth2002riemannian,
  title={Riemannian manifolds with uniformly bounded eigenfunctions},
  author={Toth, John A and Zelditch, Steve},
  journal={Duke Mathematical Journal},
  volume={111},
  number={1},
  pages={97--132},
  year={2002},
  publisher={Duke University Press}
}

@book{canuto2007spectral,
  title={Spectral methods: fundamentals in single domains},
  author={Canuto, Claudio and Hussaini, M Yousuff and Quarteroni, Alfio and Zang, Thomas A},
  year={2007},
  publisher={Springer Science \& Business Media}
}

@book{kopriva2009implementing,
  title={Implementing spectral methods for partial differential equations: Algorithms for scientists and engineers},
  author={Kopriva, David A},
  year={2009},
  publisher={Springer Science \& Business Media}
}

@book{hesthaven2007spectral,
  title={Spectral methods for time-dependent problems},
  author={Hesthaven, Jan S and Gottlieb, Sigal and Gottlieb, David},
  volume={21},
  year={2007},
  publisher={Cambridge University Press}
}

@book{shen2011spectral,
  title={Spectral methods: algorithms, analysis and applications},
  author={Shen, Jie and Tang, Tao and Wang, Li-Lian},
  volume={41},
  year={2011},
  publisher={Springer Science \& Business Media}
}

@article{meuris2023machine,
  title={Machine-learning-based spectral methods for partial differential equations},
  author={Meuris, Brek and Qadeer, Saad and Stinis, Panos},
  journal={Scientific Reports},
  volume={13},
  number={1},
  pages={1739},
  year={2023},
  publisher={Nature Publishing Group UK London}
}

@article{platte2004computing,
  title={Computing eigenmodes of elliptic operators using radial basis functions},
  author={Platte, Rodrigo B and Driscoll, Tobin A},
  journal={Computers \& mathematics with applications},
  volume={48},
  number={3-4},
  pages={561--576},
  year={2004},
  publisher={Elsevier}
}

@book{palacios2022modeling,
  author="Palacios, Antonio",
  title="Mathematical modeling: A dynamical systems approach to analyze practical problems in STEM disciplines",
  year="2022",
  publisher="Springer International Publishing"
}

@book{bennett2013transport,
  author={Bennett, Ted D},
  title={Transport by advection and diffusion: momentum, heat, and mass transfer},
  year={2013},
  publisher={Wiley}
}

@article{chaturantabut2010nonlinear,
  title={Nonlinear model reduction via discrete empirical interpolation},
  author={Chaturantabut, Saifon and Sorensen, Danny C},
  journal={SIAM Journal on Scientific Computing},
  volume={32},
  number={5},
  pages={2737--2764},
  year={2010},
  publisher={SIAM}
}

@article{ghattas2021learning,
  title={Learning physics-based models from data: perspectives from inverse problems and model reduction},
  author={Ghattas, Omar and Willcox, Karen},
  journal={Acta Numerica},
  volume={30},
  pages={445--554},
  year={2021},
  publisher={Cambridge University Press}
}

@article{karniadakis2021physics,
  title={Physics-informed machine learning},
  author={Karniadakis, George Em and Kevrekidis, Ioannis G and Lu, Lu and Perdikaris, Paris and Wang, Sifan and Yang, Liu},
  journal={Nature Reviews Physics},
  volume={3},
  number={6},
  pages={422--440},
  year={2021},
  publisher={Nature Publishing Group}
}

@article{meng2022physics,
  title={When physics meets machine learning: A survey of physics-informed machine learning},
  author={Meng, Chuizheng and Seo, Sungyong and Cao, Defu and Griesemer, Sam and Liu, Yan},
  journal={arXiv preprint arXiv:2203.16797},
  year={2022}
}

@article{hao2022physics,
  title={Physics-informed machine learning: A survey on problems, methods and applications},
  author={Hao, Zhongkai and Liu, Songming and Zhang, Yichi and Ying, Chengyang and Feng, Yao and Su, Hang and Zhu, Jun},
  journal={arXiv preprint arXiv:2211.08064},
  year={2022}
}

@article{willard2022integrating,
  title={Integrating scientific knowledge with machine learning for engineering and environmental systems},
  author={Willard, Jared and Jia, Xiaowei and Xu, Shaoming and Steinbach, Michael and Kumar, Vipin},
  journal={ACM Computing Surveys},
  volume={55},
  number={4},
  pages={1--37},
  year={2022},
  publisher={ACM New York, NY}
}

@article{yu2024learning,
  title={Learning dynamical systems from data: An introduction to physics-guided deep learning},
  author={Yu, Rose and Wang, Rui},
  journal={Proceedings of the National Academy of Sciences},
  volume={121},
  number={27},
  pages={e2311808121},
  year={2024},
  publisher={National Academy of Sciences}
}

@article{brunton2024promising,
  title={Promising directions of machine learning for partial differential equations},
  author={Brunton, Steven L and Kutz, J Nathan},
  journal={Nature Computational Science},
  pages={1--12},
  year={2024},
  publisher={Nature Publishing Group US New York}
}

@article{rao2023encoding,
  title={Encoding physics to learn reaction--diffusion processes},
  author={Rao, Chengping and Ren, Pu and Wang, Qi and Buyukozturk, Oral and Sun, Hao and Liu, Yang},
  journal={Nature Machine Intelligence},
  volume={5},
  number={7},
  pages={765--779},
  year={2023},
  publisher={Nature Publishing Group UK London}
}

@article{ren2023physr,
  title={PhySR: Physics-informed deep super-resolution for spatiotemporal data},
  author={Ren, Pu and Rao, Chengping and Liu, Yang and Ma, Zihan and Wang, Qi and Wang, Jian-Xun and Sun, Hao},
  journal={Journal of Computational Physics},
  volume={492},
  pages={112438},
  year={2023},
  publisher={Elsevier}
}

@article{faroughi2022physics,
  title={Physics-guided, physics-informed, and physics-encoded neural networks in scientific computing},
  author={Faroughi, Salah A and Pawar, Nikhil and Fernandes, Celio and Raissi, Maziar and Das, Subasish and Kalantari, Nima K and Mahjour, Seyed Kourosh},
  journal={arXiv preprint arXiv:2211.07377},
  year={2022}
}

@article{xun2013parameter,
  title={Parameter estimation of partial differential equation models},
  author={Xun, Xiaolei and Cao, Jiguo and Mallick, Bani and Maity, Arnab and Carroll, Raymond J},
  journal={Journal of the American Statistical Association},
  volume={108},
  number={503},
  pages={1009--1020},
  year={2013},
  publisher={Taylor \& Francis}
}

@article{azzimonti2015blood,
  title={Blood flow velocity field estimation via spatial regression with PDE penalization},
  author={Azzimonti, Laura and Sangalli, Laura M and Secchi, Piercesare and Domanin, Maurizio and Nobile, Fabio},
  journal={Journal of the American Statistical Association},
  volume={110},
  number={511},
  pages={1057--1071},
  year={2015},
  publisher={Taylor \& Francis}
}

@article{arnone2022some,
  title={Some first results on the consistency of spatial regression with partial differential equation regularization},
  author={Arnone, Eleonora and Kneip, Alois and Nobile, Fabio and Sangalli, Laura M},
  journal={Statistica Sinica},
  volume={32},
  number={1},
  pages={209--238},
  year={2022},
  publisher={JSTOR}
}

@article{sangalli2021spatial,
  title={Spatial regression with partial differential equation regularisation},
  author={Sangalli, Laura M},
  journal={International Statistical Review},
  volume={89},
  number={3},
  pages={505--531},
  year={2021},
  publisher={Wiley Online Library}
}

@article{arnone2023analyzing,
  title={Analyzing data in complicated 3D domains: Smoothing, semiparametric regression, and functional principal component analysis},
  author={Arnone, Eleonora and Negri, Luca and Panzica, Ferruccio and Sangalli, Laura M},
  journal={Biometrics},
  volume={79},
  number={4},
  pages={3510--3521},
  year={2023},
  publisher={Wiley Online Library}
}

@article{bernardi2017penalized,
  title={A penalized regression model for spatial functional data with application to the analysis of the production of waste in Venice province},
  author={Bernardi, Mara S and Sangalli, Laura M and Mazza, Gabriele and Ramsay, James O},
  journal={Stochastic environmental research and risk assessment},
  volume={31},
  pages={23--38},
  year={2017},
  publisher={Springer}
}

@article{arnone2019modeling,
  title={Modeling spatially dependent functional data via regression with differential regularization},
  author={Arnone, Eleonora and Azzimonti, Laura and Nobile, Fabio and Sangalli, Laura M},
  journal={Journal of Multivariate Analysis},
  volume={170},
  pages={275--295},
  year={2019},
  publisher={Elsevier}
}

@book{cressie2015statistics,
  title={Statistics for spatial data},
  author={Cressie, Noel},
  year={2015},
  publisher={John Wiley \& Sons}
}

@article{chen2022apik,
  title={APIK: Active physics-informed kriging model with partial differential equations},
  author={Chen, Jialei and Chen, Zhehui and Zhang, Chuck and Jeff Wu, CF},
  journal={SIAM/ASA Journal on Uncertainty Quantification},
  volume={10},
  number={1},
  pages={481--506},
  year={2022},
  publisher={SIAM}
}

@article{peli2022physics,
  title={Physics-based Residual Kriging for dynamically evolving functional random fields},
  author={Peli, Riccardo and Menafoglio, Alessandra and Cervino, Marianna and Dovera, Laura and Secchi, Piercesare},
  journal={Stochastic Environmental Research and Risk Assessment},
  volume={36},
  number={10},
  pages={3063--3080},
  year={2022},
  publisher={Springer}
}

@book{ramsay2005functional,
  title={Functional data analysis},
  author={Ramsay, J and Silverman, B},
  journal={New York},
  year={2005},
  edition={2},
  publisher={Springer}
}

@article{hall2007methodology,
  title={Methodology and convergence rates for functional linear regression},
  author={Hall, Peter and Horowitz, Joel L},
  journal={Annals of Statistics},
  volume={35},
  number={1},
  pages={70--91},
  year={2007},
  publisher={Institute of Mathematical Statistics}
}

@book{nickl2023bayesian,
  title={Bayesian non-linear statistical inverse problems},
  author={Nickl, Richard},
  year={2023},
  publisher={EMS press}
}

@article{nickl2020convergence,
  title={Convergence rates for penalized least squares estimators in PDE constrained regression problems},
  author={Nickl, Richard and van de Geer, Sara and Wang, Sven},
  journal={SIAM/ASA Journal on Uncertainty Quantification},
  volume={8},
  number={1},
  pages={374--413},
  year={2020},
  publisher={SIAM}
}

@article{knapik2011bayesian,
  title={Bayesian inverse problems with Gaussian priors},
  author={Knapik, BT and van der Vaart, AW and van Zanten, JH},
  journal={The Annals of Statistics},
  volume={39},
  number={5},
  pages={2626--2657},
  year={2011},
  publisher={Institute of Mathematical Statistics}
}

@article{bissantz2007convergence,
  title={Convergence rates of general regularization methods for statistical inverse problems and applications},
  author={Bissantz, Nicolai and Hohage, Thorsten and Munk, Axel and Ruymgaart, Frits},
  journal={SIAM Journal on Numerical Analysis},
  volume={45},
  number={6},
  pages={2610--2636},
  year={2007},
  publisher={SIAM}
}

@article{loubes2008adaptive,
  title={Adaptive complexity regularization for linear inverse problems},
  author={Loubes, Jean-Michel and Ludena, Carenne},
  journal={Electronic Journal of Statistics},
  volume={2},
  pages={661--677},
  year={2008},
  publisher={Citeseer}
}

@article{cavalier2008nonparametric,
  title={Nonparametric statistical inverse problems},
  author={Cavalier, Laurent},
  journal={Inverse Problems},
  volume={24},
  number={3},
  pages={034004},
  year={2008},
  publisher={IOP Publishing}
}

@article{evans2002inverse,
  title={Inverse problems as statistics},
  author={Evans, Steven N and Stark, Philip B},
  journal={Inverse problems},
  volume={18},
  number={4},
  pages={R55},
  year={2002},
  publisher={IOP Publishing}
}

@article{bissantz2008statistical,
  title={Statistical inference for inverse problems},
  author={Bissantz, Nicolai and Holzmann, Hajo},
  journal={Inverse Problems},
  volume={24},
  number={3},
  pages={034009},
  year={2008},
  publisher={IOP Publishing}
}

@article{carasso1978digital,
  title={Digital removal of random media image degradations by solving the diffusion equation backwards in time},
  author={Carasso, Alfred S and Sanderson, James G and Hyman, James M},
  journal={SIAM Journal on Numerical Analysis},
  volume={15},
  number={2},
  pages={344--367},
  year={1978},
  publisher={SIAM}
}

@article{skaggs1995recovering,
  title={Recovering the history of a groundwater contaminant plume: Method of quasi-reversibility},
  author={Skaggs, Todd H and Kabala, ZJ},
  journal={Water Resources Research},
  volume={31},
  number={11},
  pages={2669--2673},
  year={1995},
  publisher={Wiley Online Library}
}

@article{larwa2018heat,
  title={Heat transfer model to predict temperature distribution in the ground},
  author={Larwa, Barbara},
  journal={Energies},
  volume={12},
  number={1},
  pages={25},
  year={2018},
  publisher={MDPI}
}

@article{muniz1999comparison,
  title={A comparison of some inverse methods for estimating the initial condition of the heat equation},
  author={Muniz, Wagner Barbosa and de Campos Velho, Haroldo F and Ramos, Fernando Manuel},
  journal={Journal of Computational and Applied Mathematics},
  volume={103},
  number={1},
  pages={145--163},
  year={1999},
  publisher={Elsevier}
}

@article{christensen2018final,
  title={Final value problems for parabolic differential equations and their well-posedness},
  author={Christensen, Ann-Eva and Johnsen, Jon},
  journal={Axioms},
  volume={7},
  number={2},
  pages={31},
  year={2018},
  publisher={MDPI}
}

@article{klibanov2006estimates,
  title={Estimates of initial conditions of parabolic equations and inequalities via lateral Cauchy data},
  author={Klibanov, Michael V},
  journal={Inverse problems},
  volume={22},
  number={2},
  pages={495},
  year={2006},
  publisher={IOP Publishing}
}

@article{li2020recovering,
  title={Recovering the initial condition of parabolic equations from lateral Cauchy data via the quasi-reversibility method},
  author={Li, Qitong and Nguyen, Loc Hoang},
  journal={Inverse Problems in Science and Engineering},
  volume={28},
  number={4},
  pages={580--598},
  year={2020},
  publisher={Taylor \& Francis}
}

@article{minh2018two,
  title={A two-dimensional backward heat problem with statistical discrete data},
  author={Minh, Nguyen Dang and To Duc, Khanh and Tuan, Nguyen Huy and Trong, Dang Duc},
  journal={Journal of Inverse and Ill-posed Problems},
  volume={26},
  number={1},
  pages={13--31},
  year={2018}
}

@article{phuong2019cauchy,
  title={On Cauchy problem for nonlinear fractional differential equation with random discrete data},
  author={Phuong, Nguyen Duc and Tuan, Nguyen Huy and Baleanu, Dumitru and Ngoc, Tran Bao},
  journal={Applied Mathematics and Computation},
  volume={362},
  pages={124458},
  year={2019},
  publisher={Elsevier}
}

@book{hasanouglu2021introduction,
  title={Introduction to inverse problems for differential equations},
  author={Hasano{\u{g}}lu, Alemdar Hasanov and Romanov, Vladimir G},
  year={2021},
  publisher={Springer}
}

@book{isakov2017inverse,
  author="Isakov, Victor",
  title="Inverse problems for partial differential equations",
  year="2017",
  publisher="Springer International Publishing"
}
\end{document}


\def\spacingset#1{\renewcommand{\baselinestretch}%
{#1}\small\normalsize} \spacingset{1}


  \bigskip
  \bigskip
  \bigskip
\begin{center}
    {\LARGE\bf Supplementary Material for \\ ``Physics-Encoded Spatio-Temporal Regression''}
\end{center}
  \medskip

\spacingset{1.9} 

\section{Proofs}

\subsection*{Proof of Proposition~\ref{prop:evo}}
By definition, $\mathscr{M}^{-1}(\hat{u}-u)(x,t) = \sum_{k=1}^{\infty} (\hat{\alpha}_{k}-\alpha_{k}) \mathrm{e}^{-\lambda_{k}t} \phi_{k}(x)$, and thus 
\[\begin{aligned}
\norm{(\hat{u}-u)(\bm{\cdot},t)}^{2} 
&= \sum_{k=1}^{\infty} (\hat{\alpha}_{k}-\alpha_{k})^{2} \mathrm{e}^{-2\lambda_{k}t} \\
&\le \sum_{k=1}^{\infty} (\hat{\alpha}_{k}-\alpha_{k})^{2} \mathrm{e}^{-2(\inf_{k}\lambda_{k})t} \\
&= \norm{(\hat{u}-u)(\bm{\cdot},0)}^{2} \mathrm{e}^{-2(\inf_{k}\lambda_{k})t} .
\end{aligned}\]
Taking the square root yields the desired result. \qed

\subsection*{Proof of Theorem~\ref{thm:rate}}
Clearly $\hat{g}-g_{0} = \sum_{k\le K}(\hat{\alpha}_{k}-\alpha_{k})\psi_{k} + \sum_{k>K}\alpha_{k}\psi_{k}$, so by \ref{asm:coef} we have 
\[ \norm{\hat{g}-g_{0}}^{2} = \sum_{k\le K}(\hat{\alpha}_{k}-\alpha_{k})^{2} + \sum_{k>K}\alpha_{k}^{2} \lesssim \sum_{k\le K}(\hat{\alpha}_{k}-\alpha_{k})^{2} + K^{-(2s-1)} .\]
It follows from \eqref{eq:est} that 
\[ (\hat{\alpha}_{k}-\alpha_{k})_{k\le K} = (W^{\top}W)^{-1}W^{\top}(\delta+\epsilon) = (W^{\top}W)^{-1}W^{\top}(\delta+H\epsilon) ,\]
where $W = n^{-1/2}(Z_{1K},Z_{2K},\dots,Z_{nK})^{\top}$, $H = W(W^{\top}W)^{-1}W^{\top}$, $\epsilon = n^{-1/2}(\varepsilon_{1},\dots,\varepsilon_{n})^{\top}$, and $\delta = n^{-1/2}(\delta_{1},\dots,\delta_{n})^{\top}$ with 
\[ \delta_{i} = \sum_{k>K}\alpha_{k}\mathrm{e}^{-\lambda_{k}T_{i}}\psi_{k}(X_{i}) ,\quad i=1,\dots,n .\]
Denote by $\hat{\nu}_{K}$ the smallest eigenvalue of $W^{\top}W$. Since \[ (W^{\top}W)^{-1}W^{\top}\{(W^{\top}W)^{-1}W^{\top}\}^{\top} = (W^{\top}W)^{-1} ,\] the spectral norm of $(W^{\top}W)^{-1}W^{\top}$ is $\hat{\nu}_{K}^{-1/2}$. 
Then \[ \sum_{k\le K}(\hat{\alpha}_{k}-\alpha_{k})^{2} \le \hat{\nu}_{K}^{-1}(\delta+H\epsilon)^{\top}(\delta+H\epsilon) \le 2\hat{\nu}_{K}^{-1}(\delta^{\top}\delta+\epsilon^{\top}H\epsilon) .\]
Clearly $\E(\epsilon\epsilon^{\top}\mid H) = \E(\epsilon\epsilon^{\top}) = n^{-1}\Var(\varepsilon)I$, whence 
\[ \E(\epsilon^{\top}H\epsilon) = \E\{\tr(H\epsilon\epsilon^{\top})\} = \tr\{\E(H\epsilon\epsilon^{\top})\} = n^{-1}\Var(\varepsilon)\tr\{\E(H)\} = n^{-1}\Var(\varepsilon)\E\{\tr(H)\} .\]
But $H$ is a projection matrix of rank $K$, implying that $\tr(H) = K$. 
This gives \[ \epsilon^{\top}H\epsilon = \mathcal{O}_{\Pr}(n^{-1}K) .\]
It can also be seen that $\delta^{\top}\delta = \mathcal{O}_{\Pr}\{K^{-(r+2s-1)}\}$, as by \ref{asm:coef}, 
\[ \E(\delta^{\top}\delta) = \E\bigg[\Big\{\sum_{k>K}\alpha_{k}\mathrm{e}^{-\lambda_{k}T}\psi_{k}(X)\Big\}^{2}\bigg] \lesssim K^{-(r+2s-1)} .\]
Putting the above together, it remains to show that \[ \hat{\nu}_{K}^{-1} = \mathcal{O}_{\Pr}(K^{r}) .\] 
Let $\Delta = W^{\top}W - \E(Z_{K}Z_{K}^{\top})$ whose $(k,k')$th entry is denoted by $\Delta_{kk'}$. We have 
\[\begin{aligned}
\E(\norm{\Delta}_{\mathrm{F}}^{2}) &= \sum_{k,k'\le K}\E(\Delta_{kk'}^{2}) \\
&= \sum_{k,k'\le K} \Var\Big\{ n^{-1}\sum_{i\le n} \mathrm{e}^{-(\lambda_{k}+\lambda_{k'})T_{i}}\psi_{k}(X_{i})\psi_{k'}(X_{i}) \Big\} \\
&= n^{-1} \sum_{k,k'\le K} \Var\{\mathrm{e}^{-(\lambda_{k}+\lambda_{k'})T}\psi_{k}(X)\psi_{k'}(X)\} \\
&\le n^{-1} \sum_{k,k'\le K} \E\{\mathrm{e}^{-2(\lambda_{k}+\lambda_{k'})T}\psi_{k}^{2}(X)\psi_{k'}^{2}(X)\} \\
&= n^{-1} \E\bigg[\Big\{\sum_{k\le K} \mathrm{e}^{-2\lambda_{k}T} \psi_{k}^{2}(X)\Big\}^{2}\bigg] \\
&\lesssim K^{-r-2s+c} = o(K^{-2r}) 
\end{aligned}\]
by \ref{asm:cov} and that $2s>r+c$. 
The spectral norm $\norm{\Delta}$ of $\Delta$ is bounded by its Frobenius norm $\norm{\Delta}_{\mathrm{F}} = o_{\Pr}(K^{-r})$, and thus \ref{asm:eigen} leads to that 
\[ \hat{\nu}_{K} \ge \nu_{K} - \norm{\Delta} \gtrsim K^{-r}\{1-o_{\Pr}(1)\} .\]
The proof is then complete. \qed

\subsection*{Proof of Theorem~\ref{thm:minimax}}
Consider $\mathscr{L}$ and $P$ such that $\varepsilon$ follows the standard normal distribution, $X$ and $T$ are independent, $\E\{\psi_{k}(X)\psi_{k'}(X)\} = \delta_{kk'}$ and $\E(\mathrm{e}^{-2\lambda_{k}T}) \le C_{\lambda} k^{-r}$ for some constant $C_{\lambda}>0$. 
Let $K = K_{n}$ be the the integer part of $C_{0}n^{1/(r+2s)}$ where $C_{0}$ is a constant to be specified later. 
For any $\theta = (\theta_{1},\dots,\theta_{K})^{\top} \in [0,1]^{K}$, write 
\[ g_{\theta} = \sum_{K<k\le 2K} k^{-s} \theta_{k-K} \psi_{k} .\]
Given $\tilde{g} = \sum_{k=1}^{\infty} \tilde{\alpha}_{k} \psi_{k}$, define $\tilde{\theta} = (\tilde{\theta}_{1},\dots,\tilde{\theta}_{K})^{\top}$ such that $\tilde{\theta}_{k-K} = \min\{(k^{s}\tilde{\alpha}_{k})_{+},1\}$ for $K<k\le 2K$, where $a_{+}=\max\{a,0\}$. 
It is straightforward that for any $\theta\in[0,1]^{K}$, 
\[ \norm{\tilde{g}-g_{\theta}}^{2} \ge \sum_{K<k\le 2K} (\tilde{\alpha}_{k}-k^{-s}\theta_{k-K})^{2} \ge \sum_{K<k\le 2K} k^{-2s} (\tilde{\theta}_{k-K}-\theta_{k-K})^{2} \ge (2K)^{-2s} \lVert\tilde{\theta}-\theta\rVert^{2} ,\]
where by abuse of notation, $\lVert\vartheta\rVert$ denotes the Euclidean norm of $\vartheta$ for $\vartheta\in\mathbb{R}^{K}$. 
Then \[\begin{aligned}
\Pr\{ \norm{\tilde{g}-g_{0}}^{2} > \delta n^{-(2s-1)/(r+2s)} \} 
&\ge \Pr\{ \lVert\tilde{\theta}-\theta\rVert^{2} > 2^{2s}\delta K^{2s} n^{-(2s-1)/(r+2s)} \} \\
&\ge \{\E(\lVert\tilde{\theta}-\theta\rVert^{2}) - 2^{2s}\delta K^{2s} n^{-(2s-1)/(r+2s)}\}_{+}^{2} / \E(\lVert\tilde{\theta}-\theta\rVert^{4}) \\
&\ge \{\E(\lVert\tilde{\theta}-\theta\rVert^{2}) / K - 2^{2s} C_{0}^{2s-1} \delta \}_{+}^{2} , 
\end{aligned}\]
where the second step applies the Paley--Zygmund inequality \citep[Lemma~5.1]{kallenberg2021foundations}. 
By setting $\delta = (2C_{0})^{-2s}$, it suffices to show that \[ \liminf_{n\to\infty} \inf_{\tilde{\theta}} \sup_{\theta\in[0,1]^{K}} \E_{(g_{\theta},\mathscr{L},P)}(\lVert\tilde{\theta}-\theta\rVert^{2}) / K > C_{0}^{-1} .\]
Using the Varshamov--Gilbert bound \citep[Lemma~2.9]{tsybakov2009introduction}, there exists a set of binary sequences $\{\theta^{(1)},\dots,\theta^{(M)}\}\subset\{0,1\}^{K}$ such that $M \ge 2^{K/8}$ and 
\[ \lVert\theta^{(j)}-\theta^{(j')}\rVert^{2} \ge K/8 ,\quad \forall j\ne j' .\]
Denote by $P_{nj} = P_{j}^{\otimes n}$ the distribution of $\{(U_{i},X_{i},T_{i})\}_{i=1,\dots,n}$ under $(g_{\theta^{(j)}},\mathscr{L},P)$. 
Since the conditional distribution of $U_{i}\mid (X_{i},T_{i})$ is Gaussian with mean $u_{g_{0}}(X_{i},T_{i})$ and variance $1$, the Kullback--Leibler divergence of $P_{nj}$ with respect to $P_{nj'}$ is given by 
\[\begin{aligned}
\operatorname{KL}(P_{nj} \Vert P_{nj'}) 
&= n \operatorname{KL}(P_{j} \Vert P_{j'}) = n \int \log\Big(\dv{P_{j}}{P_{j'}}\Big) \dd{P_{j}} \\
&= n \, \E[\{u_{g_{\theta^{(j)}}}-u_{g_{\theta^{(j')}}}\}^{2}(X,T)] \\
&= n \sum_{K<k\le 2K} k^{-2s} (\theta_{k-K}^{(j)}-\theta_{k-K}^{(j')})^{2} \E(\mathrm{e}^{-2\lambda_{k}T}) \\
&\le C_{0}^{-(r+2s)} (K+1)^{r+2s} C_{\lambda} K^{-r-2s+1} 
\le (C_{0}/2)^{-(r+2s)} C_{\lambda} K .
\end{aligned}\]
Therefore, an application of Fano's lower bound \citep[Proposition~15.12]{wainwright2019high} gives 
\[ \inf_{\tilde{\theta}} \sup_{\theta\in[0,1]^{K}} \E_{(g_{\theta},\mathscr{L},P)}(\lVert\tilde{\theta}-\theta\rVert^{2}) \ge \frac{K}{32} \Big\{ 1 - \frac{(C_{0}/2)^{-(r+2s)} C_{\lambda} K + \log(2)}{(K/8)\log(2)} \Big\} .\]
Choosing $C_{0}$ large enough completes the proof. \qed


\bibliographystyle{agsm}
\bibliography{ref}